# Nanowires for heat conversion

Milo Yaro Swinkels[1] and Ilaria Zardo[1]

[1] Departement Physik, Universität Basel, Klingelbergstrasse 82, 4056 Basel, Switzerland

E-mail: miloyaro.swinkels@unibas.ch, ilaria.zardo@unibas.ch

**Abstract**

This review focuses on the investigation and enhancement of the thermoelectric properties of semiconducting nanowires (NWs). NWs are nanostructures with typical diameters between few to hundreds of nm and length of few to several μm, exhibiting a high surface-to-volume ratio. Nowadays an extraordinary control over their growth has been achieved, enabling also the integration of different type of heterostructures, which can lead to the engineering of the functional properties of the NWs. In this review we discuss all concepts which have been presented and achieved so far for the improvements of the thermoelectric performances of semiconducting NWs. Furthermore, we present a brief survey of the experimental methods which enable the investigation of the thermoelectric properties of these nanostructures.

Kewords: Nanowires, core-shell nanowires, superlattice nanowires, thermal conductivity, Seebeck coefficient, figure of merit ZT, thermoelectrics

## 1 Introduction

An estimated 70% of the produced energy is lost as heat. Thermoelectric devices offer the possibility to turn this heat into electricity and could thus be employed to greatly reduce the energy footprint. However, thermoelectric devices have two main drawbacks; a low efficiency due to a poor thermoelectric performance of the available materials and high cost due to scarcity of available materials. The reason for this limitation can be seen from the thermoelectric figure of merit ZT, defined as:

$$ZT = \frac{S^2 \sigma}{\kappa} T$$

where S is the Seebeck coefficient, $\sigma$ the electrical conductivity and $\kappa$ the thermal conductivity. The reason why it is challenging to find materials with a high thermoelectric figure of merit is because most of the parameters are interdependent i.e. an increase in the Seebeck coefficient usually means a decrease in electrical conductivity and an increase in electrical conductivity generally means an increase in thermal conductivity as well [1].

One way to overcome these obstacles and to find materials that enable large-scale application of thermoelectric devices is by employing nanoengineering to enhance the thermoelectric performance of even more common materials. This boom in research has been triggered by two papers by Hicks et. al. where a prediction is made that the thermoelectric figure of merit could be larger for 2D[2] and 1D[3] conductors due to modifications in the density of states. Secondly, it was predicted that by employing typical length scales smaller than the mean free path of phonons the phonon contribution to the thermal conduction, which is the biggest contribution to thermal conductivity in semiconductors, can be reduced, while keeping the electrical properties unchanged [4]. This has been shown experimentally on rough silicon[5,6], which triggered a wide range of both theoretical and experimental work on phonon transport in nanoscaled devices.

Nanowires (NWs) offer model systems for testing predictions on thermoelectric performance at the nanoscale thanks to their wide range of fabrication techniques offering a lot of freedom in material choice and geometries offering control over the surface-to-volume ratio. For example, NWs have enabled the creation of complex heterostructures[7], core-shell[8–10] and crystal phase[11,12] structures unattainable in bulk materials. Secondly, NWs approach one-dimensional transport systems, simplifying theoretical predictions. In this review paper, we discuss recent progress made on the thermoelectric properties of NWs as well as theoretical predictions that have been made. We provide first a survey of the different approaches that have been employed to enhance the ZT values in NWs, including the reduction of thermal conductivity as well as the enhancement of the power factor $S^2\sigma$. Next, the different measurement techniques used are described, including their advantages and shortcomings. Finally an overview of ZT values found and predicted will be given.

## 2 Strategies to enhance the thermoelectric efficiency of NWs

In this section, results obtained on the thermoelectric performance of NWs will be discussed. The first part is dedicated to the decrease of thermal conductivity, while the second part is about the increase of the power factor.

### 2.1 Decrease of thermal conductivity

In crystals, the thermal conductivity $\kappa$ is generally derived from a kinetic theory of the phonon gas, and it is defined as the energy transmitted per unit time across unit area per unit temperature gradient through Fourier's law: $j = -\kappa \nabla T$, where $j$ is the flux of thermal energy and $\nabla T$ is the temperature gradient. The thermal conductivity can be decomposed in the electronic and lattice contributions: $\kappa = \kappa_{electronic} + \kappa_{lattice}$. In semiconductors, the thermal conductivity is dominated by the lattice thermal transport, which can be written as $\kappa_{lattice} = \frac{1}{3} C_v v_s \lambda_{ph}$, where $C_v$ is the lattice heat capacity, $v_s$ is the sound velocity, and $\lambda_{ph} = v\tau$ is the phonon mean free path, with $\tau$ phonon lifetime.

The phonon mean free path (or equivalently the phonon lifetime) is limited by boundary scattering and phonon-phonon scattering. Boundary scattering of phonons in a nanostructure is expected to lower $\lambda_{ph}$, with a consequent reduction of the thermal conductivity. Alloy point defect scattering (*e.g.* $Si_{1-x}Ge_x$) also lead to the reduction of $\lambda_{ph}$. On the other hand, heavy element compounds semiconductors (*e.g.* $Bi_2Te_3$) have low sound velocities and, therefore, low thermal conductivities. Complex crystal structures also lead to low thermal conductivity due to a significantly reduced population of acoustic phonons that dominate the lattice thermal conductivity since they have a higher velocity with respect to optical phonons. Concerning the temperature dependence of the thermal conductivity, we have to consider the temperature dependence of $C_v$ and $\lambda_{ph}$. At high temperature, $\lambda_{ph} \propto 1/T$ since the probability of phonon-phonon scattering increases with the phonon population. On the other hand, at very low temperatures, the thermal conductivity is limited by the surface scattering and the temperature dependence of $\kappa$ is attributed entirely to the $T^3$-dependence of the specific heat. The thermal conductivity reaches its peak value when the length of the mean free path due to Umklapp-processes is comparable to the mean free path of the surface scattering. For higher temperatures, the thermal conductivity decreases since the rate of Umklapp-scattering increases homogeneously.

Therefore, phonon scattering at boundaries and interfaces is an effective method to reduce the thermal conductivity $\kappa$ of materials. Since for thermoelectrics a reduction in thermal conductivity is desired while a reduction in electrical conductivity is detrimental, a material with a high phonon mean free path but a small electron mean free path is optimum for using this method to increase ZT. The exact dependency of the thermal conductivity on the diameter (D) depends on the bulk mean free path of the material and can thus be very different for different materials. Interestingly, D. Donadio and G. Galli found that the size reduction and phonon confinement do not necessarily lead to low values of the thermal conductivity [13]. Furthermore, because not all phonons will undergo diffusive scattering on the boundary the exact shape of the boundary will have a strong influence as well. In this respect, surface roughness is a particularly important ingredient [14]. Nevertheless, the thermal conductivity can be successfully engineered only when both core defects and surface structure and composition are tailored, as shown in the study of Y. He and G. Galli [15]. In this section, these different dependencies and how they have been measured will be discussed.

### 2.1.1 Boundary scattering

When the diameter of a nanowire is reduced below the phonon mean free path, the thermal conductivity is reduced [16–18]. As a first approximation Matthiessen's rule allows for the calculation of the mean free path reduction as a result of nanostructuring by adding the contributions of impurity scattering, Umklapp scattering and boundary scattering together. However, for more rigorous calculations the mean free path as a function of vibrational frequency must be considered. Additionally, diffuse scattering of phonons is enhanced further in rough NWs, reducing the thermal conductivity. This effect depends on the root-mean-square (RMS) value of the roughness as well as the autocovariance length, which can be tuned to favour scattering in a certain frequency range. Martin et. al. proposed a model where for thin NWs, the thermal conductivity scales as $(D/\Delta)^2$ where $\Delta$ is the RMS roughness of the sample. This model has been used to successfully explain the thermal conductivity of rough silicon [19] as well as Ge and GaAs NWs[20]. Figure 1(a) and (b) compare the experimental results on the temperature dependence of the thermal conductivity for smooth (a) and rough (b) silicon NWs of different dimensions. Figure 1(c) shows the model proposed by Martin et. al. showing the dependence of the thermal conductivity on the surface roughness for two different diameters for several different materials, showing the importance of surface roughness.

In the following, we will review the main results achieved in the study of the effect of boundary scattering on the thermal conductivity of elementary and binary compound NWs.

For elementary wires two material systems have been extensively studied: bismuth and silicon. Silicon nanowires are by far the most studied elementary nanowires. This due to the extensive knowledge in silicon processing as well as the abundance of the material, making it an ideal candidate for cheap thermoelectric applications. Measurements by Hochbaum et. al[5] and Boukai et. al. [6] both showed the promise of silicon nanowires as thermoelectric materials by showing a reduction of the thermal conductivity by up to two orders of magnitude while maintaining electrical performance. Figure 2 a-c show the measured electrical and thermal conductance as well as the power factor measured in the work of Hochbaum *et al.*. Later theoretical work was able to reproduce the result using frequency dependent boundary scattering[19]. The importance of surface roughness and the resulting frequency dependent scattering has been further established in the experimental work of Lim et. al.[21], where the thermal conductivity of silicon NWs with different roughness was measured.

Figure 2 d-f show the comparison of the experimental data for the dependence of $\kappa$ on root mean square (rms) $\sigma$, correlation length L, which is a statistical parameter that determines the decay of the autocovariance and is related to the lateral length scale of the roughness, and high-frequency amplitude term $\alpha_p$, which is a parameter that increases with surface roughness. From this work it was concluded that surface roughness causes a frequency-dependent scattering of phonon modes, which depends strongly on the roughness parameters and will, in certain conditions, have a greater effect than the reduced diameter. Chen et. al. showed that a combination of specular and diffusive scattering, the ratio of which is given by the specularity parameter, in silicon NWs with a diameter of less than 30 nm causes a linear temperature dependence[22]. This work was modelled theoretically using a Landauer formalism as established in [23]. Finally, Moore et. al. showed that the specific shape of the surface could even be important by theoretically modelling phonon backscattering in sawtooth NWs, an effect that would otherwise only be accounted for by using a negative specularity parameter[24]. Finally, theoretical work from F. Sansoz showed that the thermal conductivity also strongly depends on the surface faceting in Silicon NWs[25] showing once again the importance of the surface in these systems.

Bismuth NWs have been studied due to their wide application as a bulk material for thermoelectrics as early as 1999[26]. By using the nanowire morphology, the main increase in ZT value is predicted to come from the decrease of thermal conductivity, resulting in ZT values up to 4 NWs with a diameter below 10 nm[27]. Experimental results show a reduction of 18-78 times of the thermal conductivity for polycrystalline Bi nanowires with a diameter of 74-255 nm, down to a value of approximately 0.7 Wm$^{-1}$K$^{-1}$[28]. Interestingly, Roh *et. al.* found a strong dependence on the growth direction of the thermal conductivity of Bi nanowires, consistent with the anisotropy in the thermal conductivity of bulk bismuth. Namely, Bi nanowires with a diameter of 100 nm grown in the [$\bar{1}$02] direction showed a thermal conductivity 7 Wm$^{-1}$K$^{-1}$ higher than wires grown in the [110] direction[29]. Figure 3(a) shows the diameter-dependent thermal conductivity of Bi nanowires for the two different growth directions, while 3(b) shows the selected area electron diffraction (SAED) measurements showing the directionality of the wires. Figure 3(c) shows a theoretical prediction for the ZT for Bi NWs as a function of diameter and direction.

Of III-V nanowires, InSb and InAs have been predicted to show the largest potential for thermoelectric devices, with expected ZT values up to 5 for NWs with a diameter below 20 nm[30]. Figure 4(a) and (b) show the predicted lattice thermal conductivity and ZT for a wide range of III-V nanosystems as a function of typical length scale. This increase was predicted to be mainly coming from a decrease in thermal conductivity. A study on InSb nanowires showed a decrease in thermal conductivity for wires with a diameter of approximately 150 nm[31], however a different study found InSb nanowires with this diameter to have a thermal conductivity comparable to bulk[32]. The discrepancy between these two results has been explained by a difference in crystalline quality between the wires. A small reduction of the thermal conductivity of InSb for this diameter is also expected, taking into account the mean free path of phonons in InSb which is estimated to be around 84 nm[33]. A measurement on thinner InSb nanowires has not been performed yet. Figure 4(c) shows a comparison between experimental results and theoretical models for the temperature dependence of the thermal conductivity of InSb NWs

As InSb, InAs is also a promising thermoelectric material as well, thanks to excellent electrical properties such as low electron effective mass and high electron mobility. Furthermore, InAs offers a relative ease in the fabrication of contacts, thanks to Fermi band pinning. Several studies have been performed on the thermal conductivity of InAs NWs[34–36] all showing a reduction of the thermal conductivity to below 10 Wm$^{-1}$K$^{-1}$ for a diameters smaller than 100 nm. For wires with a diameter of 40 nm a reduction of the thermal conductivity of 80% was found[36]. However, simultaneous measurement of the electrical characteristics showed a ZT value of $5.4 \times 10^{-4}$ due to reduced electrical performance[35], which was attributed to surface charges and defects.

Bulk bismuth telluride ($Bi_2Te_3$) has for a long time been considered one of the most efficient thermoelectric materials available, thanks to a zT~1 near room temperature. Therefore, significant research effort has been devoted to measure the influence of nanostructuring on $Bi_2Te_3$. However, a full thermoelectric characterization of $Bi_2Te_3$ found no reduction of the thermal conductivity for wires with a diameter of approximately 120 nm[37]. Another study found a 20% reduction of the thermal conductivity for wires with a diameter of 52-55 nm[38], however neither of the papers reported an increase of ZT above the bulk value. In $Bi_2Te_3$ the electron mean free path has been estimated to be 61 nm at room temperature[38], while the phonon mean free path has been estimated to be around 3 nm[38]. This means that, in the case $Bi_2Te_3$, nanostructuring will reduce electrical properties before reducing the thermal conductivity, exactly the opposite of what is desired for thermoelectrics.

Lead based materials such as lead telluride (PbTe) and lead selenide (PbSe) suffer from the fact that lead is a toxic element, and thus safely incorporating these materials into thermoelectric devices presents a significant challenge. The thermal conductivity of PbTe NWs has been measured for diameters down to 180 nm, for which a thermal conductivity of approximately 1.3 Wm$^{-1}$K$^{-1}$ at room temperature was found (*i.e.* a reduction of 50% with respect to bulk), in two independent studies[39,40]. A molecular dynamics combined with Boltzmann study found that a significant reduction of the thermal conductivity of PbTe with respect to bulk is only expected for samples with a feature size around or below 10 nm[41]. Furthermore, both the theoretical and the experimental work show a further decrease of the mean free path with temperature, causing a further decrease of the effect of the reduced feature size on the thermal conductivity of PbTe NWs at higher temperatures[40,41].

Similar to PbTe, PbSe has been predicted to require similar feature sizes of approximately 10nm to achieve significant reduction of the thermal conductivity[42]. Experimental results on NWs with a diameter of 180 nm found a 10-1000 times reduction of the thermal conductivity with respect to bulk below 100K, but this reduction vanishes around room temperature[43]. Another study found a two-fold reduction of the thermal conductivity to 0.8 Wm$^{-1}$K$^{-1}$ for PbSe NWs with a diameter of 50-100 nm[44] also reporting a sharp drop of the thermal conductivity below 100K.

*2.1.2 Alloys*

Apart from boundary scattering, other scattering mechanisms can also be enhanced in NWs in order to further reduce the thermal conductivity and improve thermoelectric performance. Another possible scattering centre for phonons is a mass defect, *i.e.* a substitutional impurity atom with different mass like for example impurities or alloying. Although impurities generally reduce electrical performance as well, alloying can have a relatively small effect on electronic performance, while greatly reducing thermal transport. Since alloy scattering mainly affects low-wavelength phonons, while boundary scattering is frequency independent above a certain cut-off frequency[23,45], combining these two scattering mechanisms can lead to a further reduction of the thermal conductivity. The alloy scattering rate of alloy scattering is proportional to $\sqrt{M_1/M_2}$ so a larger mass mismatch leads to an increased alloy scattering rate. The most common alloy studied and used for thermoelectric applications is SiGe, which has been used for example in NASA's deep space missions for thermoelectric power generation[46].

The fundamental lower limit to the thermal conductivity of alloys caused by alloy scattering is the alloy limit. However, for SiGe there is some debate about what this limit exactly is, and theoretical and experimental work show a disagreement on the exact value of the alloy limit[47,48]. It has been shown that the amount of alloy scattering depends strongly on the amount of disorder of the atoms in the alloy[47]. Lee et. al. showed studied the effect of the size of a supercell with a random alloy distribution. By increasing the size of the supercell the random distribution of alloy atoms could be further randomized, leading to a further reduction in the thermal conductivity[48]. A minimum thermal conductivity of 1.53 Wm$^{-1}$K$^{-1}$ was found for bulk SiGe for the largest disorder in atom structure [48].

Rather than increasing randomness, the creation extended islands of material insertions have also been shown to enable reduction of the thermal conductivity, even below the amorphous value[49]. This was explained by the fact that there is a relation between the size of these grains and the wavelength of phonons they mainly scatter. By tuning the island size and distribution this can be used to scatter a wider range of phonon frequencies, further reducing the thermal conductivity. Furthermore, it was shown that for nanocomposites, the thermal conductivity is more influenced by the size distribution of the grains than by the stochiometry[50].

In 2010, Wang and Mingo theoretically computed the thermal conductivity of SiGe alloy NWs as a function of NW diameter, alloy concentration, and temperature [51]. Their results are reported in Figure 5. Interestingly, they found a weaker diameter dependence beyond a certain onset diameter, which depends on the alloy concentration. These findings exposed the coexistence of alloy and boundary scattering contributions. Measurements of the thermal conductivity of SiGe NWs provided a value for the thermal conductivity of 1.1 Wm$^{-1}$K$^{-1}$ for tapered NWs with a mean diameter of about 200 nm[52]. Another work showed a thermal conductivity of 1.2 Wm$^{-1}$K$^{-1}$ for a NW with a diameter of 65 nm[53]. The conductivity was found to show little dependence on the diameter, but a large dependence on the alloy distribution. These results show that boundary scattering has a relatively small effect on SiGe NWs.

However, both theoretical[54] and experimental[55] showed that the roughness of the sidewall can still have a considerable effect, although smaller than the effect on silicon NWs. This can be explained that in alloys in general low-frequency phonons, which are less influenced by the boundary, are the dominant energy carriers. In fact, SiGe NWs have also shown room temperature ballistic phonon transport over 8.3 µm[56], which was also explained by the filtering of low mean free path, high frequency phonons due to alloying.

A special type of mass mismatch scattering can be achieved using isotopes. Since isotopes have exactly the same electronic properties, these are not altered when adding different isotopes of a material. However, the different masses of the isotopes will cause an increase in phonon scattering. Wei et. al. showed that the thermal conductivity of isotopically purified diamond is up to 4 times higher than that of diamond with the natural isotope distribution[57]. As with alloy scattering, theoretical work on silicon nanowires showed that even a small percentage of isotope impurity can already have a large effect on the thermal conductivity[58–60]. To date, only one experimental work has been published on the effect of isotopes in NWs. Here a mixed $^{28}Si_x^{30}Si_{1-x}$ nanowire was compared with an isotopically purified $^{29}Si$ nanowire and a reduction of 30% of the thermal conductivity of the isotopically mixed nanowire was found. This work had an unintentional radial distribution of isotopes where the concentration of heavier isotopes increased towards the center. However, later theoretical work showed that such a radial distribution has little effect on the thermal conductivity[60].

### 2.1.3 Superlattices

Recently there has been a lot of interest in using superlattices to further reduce the thermal conductivity of NWs. Where alloying inherently has a length scale on the scale of the lattice parameter and thus mainly targets the same high frequency phonons suppressed with boundary scattering, a superlattice structure allows to tune the length scale and with that also the phonons targeted for increased scattering. This allows for the suppression of thermal transport from a wider range of phonons and thus can lead to a further reduction of the thermal conductivity, especially when combined in NWs, where the high frequency phonons are already scattered through boundary scattering. Dames and Chen showed that combining superlattices with the nanowire geometry could lead to a factor of 2 reduction of the thermal conductivity with respect to conventional superlattices[61]. It has been predicted that by designing a superlattice with suitable period the thermal conductivity can even be reduced below the amorphous limit[62], which was for a long time considered the lowest limit for the thermal conductivity of a material. Another advantage for implementing these superlattices in NWs is that it allows for a wider range of materials to be combined in axial superlattices thanks to the relaxation of lattice stress in the lateral direction[7]. Finally, superlattices allow not only for a surpression of the thermal conductivity, but also for an increase of the power factor, it is discussed in more detail in Section 2.2[63].

Superlattice structures can be formed using a wide variety of periodic structures including a periodicity in material system[61,63–66], geometry[67,68] or crystal structure[69,70]. A Non-Equilibrium Molecular Dynamics (NEMD) study of Si/Ge superlattice NWs showed that the thermal conductivity of superlattice NWs could be almost as low as the alloy counterpart[65], as is also shown in figure 6(a) where the thermal conductivity of a crystalline NW, a superlattice NW, an alloy NW and an amorphous NW are compared. Furthermore, it was found that an optimum exists for the period length, due to a competition between interface resistance reducing phonon group velocities and coherent phonons occurring at very short period lengths that facilitate thermal transport. This effect was also used to propose NWs with non-uniform barrier widths to filter a wider range of phonon frequencies[71]. Finally, apart from changing the material, NWs also opened up the possibility to change the crystal structure from a cubic to a hexagonal structure in a controllable way[11]. Indeed, density functional theory (DFT) and Boltzmann transport equation (BTE) calculations have shown that the thermoelectric properties of Si and Ge NWs depend on the type of polytype [72]A theoretical work on SiC showed that such a switching of crystal structure could also further reduce the thermal conductivity[69], the results of this are shown in figure 6(c) where the thermal conductivity in a uniform NW is compared with the thermal conductivity of SL NWs formed either by periodically changing the diameter or the crystal structure. Another example of this was obtained by comparing the thermal conductivity of NWs with a twinning superlattice[70], as shown in figure 6(d). Here an optimum period spacing that depends on the diameter of the NW was found to achieve the maximum reduction of the thermal conductivity.

Superlattices can also be created by periodically varying the geometry. In the case of NWs, this is usually done by changing the diameter. It has been shown that by periodically changing the diameter of Si and Si/$SiO_2$ cross-section-modulated nanowires phonons can be trapped inside modulated nanowire segment, reducing the thermal conductivity by a factor 3-7[67]. This result has been extended in a later theoretical work where the effect of a diameter modulated amorphous shell around a crystalline core was investigated. It was found that by using a diameter modulated amorphous shell the thermal conductivity could be reduced to nearly the value of a completely amorphous wire[68], as shown in figure 6(b) while the existence of a crystalline core would mean much better electrical properties.

To the best of our knowledge, the only experimental work performed on the thermal conductivity of superlattice nanowires has been performed on Si/SiGe superlattices[73]. In this work it was found that the main reduction of the thermal conductivity was coming from alloy scattering in the SiGe sections of the nanowire, with a contribution from boundary scattering as well. Interface scattering was most likely low due to Ge diffusion into the Si segments reducing the interface sharpness. Furthermore, the acoustic mismatch between the Si and the SiGe segments was relatively small due to the low concentration of germanium of 5-10%, further reducing the effect of interface scattering. Studies on 2D Si Si/Ge superlattices have shown that a percentage difference of 30% is required to have a significant acoustic mismatch and thus to cause phonon reflection at the interfaces and typical superlattice effects[74–76].

### 2.1.4 Core-shell structures

Due to the fact that in typical NWs the diameter is a lot smaller than the mean free path, phonons in this system are strongly influenced by the surface. Core-shell structures in NWs attempt to employ this surface sensitivity by engineering the surface in a way to further reduce the thermal conductivity. The first theoretical proposal for enhanced thermoelectric performance in core/shell NWs came from Yang et. al. [77]. Since then several other theoretical studies have been performed to study the effect of the core/shell structure on the thermoelectric properties[78–84] A molecular dynamics study on the effect of an atomistic Si-coating on a Ge nanowire with a diameter of 2-12 nm found that in core-shell NWs two competing mechanisms can take place. First of all the coating increases the overall diameter of the nanowire and thus reduces boundary scattering. However, the coating was also found to cause a coupling between longitudinal and transverse phonon modes, leading to a reduction of the thermal conductivity[85]. This meant that there is a critical shell thickness for which the thermal conductivity is minimal, which depends on the material in both the core and the shell, as well as the core thickness. Furthermore, the maximum decrease of the thermal conductivity achievable through atomistic coating was found to increase with increasing core diameter. This effect is also shown in figure 7(a) where the normalized thermal conductivity as a function of thickness of the coating is shown for Ge NWs with different diameters.

An experimental study on the thermal conductivity of Bi/Te core-shell nanowires with a diameter of 160-200 nm found a reduction of the thermal conductivity by a factor 2 compared to pure Bi nanowires with a similar diameter, which could be reduced further by creating a rough interface between the bismuth and telluride, causing another two to five-fold reduction[86]. Similar conclusions were drawn in a theoretical study by Wu and Wu where a Boltzmann transport formalism was used to study the effect of surface roughness on the thermoelectric properties of InN/GaN core/shell NWs. It was found in this study that a root-mean-square value of the roughness of 3 nm caused a reduction of the thermal conductivity by a factor 15 as compared to a completely smooth core/shell NW [83]. Mainly thanks to this reduction in thermal conductivity an enhancement of the ZT value up to 1.713 was predicted in this paper.

In an experimental study on Si/Ge core-shell NWs with a diameter of approximately 20 nm it was found that the thermal conductivity of the core-shell NW is higher than that of a uniform wire at low temperatures, while being lower above room temperature[87]. This was explained by reduction of the thermal conductivity in core-shell NWs due to coupling between longitudinal and transverse modes, which is frequency and thus also temperature dependent. A more recent study on Bi/Te core-shell NWs with a rough interface also shows a reduction of the thermal conductivity with respect to uniform wires and also explains this with additional scattering of phonons in the rough interface region[10].

In real-world applications NWs will generally have an amorphous shell of native oxide around them. The influence of this naturally occurring shell has also been studied and it has been found that even a few nm thick amorphous shell, which is the typical thickness of a native oxide shell[88], can reduce the thermal conductivity by a factor 2[89]. This result is also shown in figure 7(b) where the thermal conductivity as a function of diameter is plotted for different core and amorphous shell thicknesses.

Apart from single core/shell structures, multi core/shell (MCS) NWs have also been studied theoretically for their influence on the thermoelectric properties. In 2014, Zhou and Chenthe used a non-equilibrium Green's function and molecular dynamics study to find the thermoelectric properties of MCS GaSb/InAs NWs [84]. A reduction of more than a factor 2 in phonon thermal conductivity at room temperature was found when going from a pure GaSb nanowire to a GaSb/InAs core/shell structure. Another factor of 2 improvement was obtained when adding more shell layers. Interestingly, this reduction in phonon thermal conductivity completely vanished at temperatures above 600 K. Since the introduction of the core/shell structure was predicted to slightly decrease electrical performance, this means that MCS NWs had the highest thermoelectric efficiency at 300 K, while above 500 K the pure NWs were predicted to have the highest ZT. This effect is also shown in figure 7(c) and (d) where the maximum ZT and the thermal conductivity of difference MCW NWs as a function of temperature is compared.

## 2.2 Increase in power factor

Apart from reducing the thermal conductivity, the thermoelectric figure of merit can also be increased by increasing the power factor $\sigma S^2$. The effect has first been proposed by Hicks and Dresselhaus in 1993 for 2D[2] and 1D[3] materials. Due to a modification of the Density of States (DOS) due to quantum confinement the DOS can become very asymmetric around the Fermi energy. A strong asymmetry around the Fermi energy causes a larger dependence on the temperature and thus a larger Seebeck coefficient, while the electrical conductivity is potentially unaffected, increasing the power factor. However, the original proposal was written for a single band system, which is valid as long as the confinement is strong enough or the electron mass small enough to sufficiently split the bands. A more elaborate model taking account multiple bands showed that when the single-band condition is not sufficiently satisfied the power factor will actually decrease with decreasing diameter, and only below a certain critical diameter depending on material parameters the power factor of a nanowire would increase[90]. This might explain why an enhancement of the thermopower in NWs has only been observed in a very small number of publications where other effects were most likely also playing a role[91,92]. One possible mechanism through which sufficient band splitting to observe an enhancement of the power factor can still be achieved in NWs is by employing superlattice structures[63], which can be designed to change the DOS through the formation of minibands.

An enhancement of the thermoelectric power factor in NWs was first reported in a study on InAs nanowires[93]. In one research InAs nanowires with a diameter of approximately 60 nm showed an improvement of the power factor at temperatures below 20 K. These wires are much thicker than the critical diameter where an enhancement of the thermopower would be expected. However, further study of the nanowires indicated that 0D transport was taking place in these wires and could thus explain of the enhancement. This 0D transport behavior was attributed to nonuniformities in the nanowire. In another report SnTe NWs with a diameter of 218-913 nm were also reported to show an increase in the power factor[94]. The enhancement was found to increase with decreasing diameter, however also in this case the diameters were far too large to attribute the enhancement to quantum confinement effects and the effect was instead attributed to uniaxial strain, that when combined with a strong spin-orbit coupling could lead to substantial changes in the energy bands. In conclusion, the proposed increase in thermoelectric power factor for NWs has so far only been achieved when coupled with other effects, such as 0D confinement or spin-orbit mediated band splitting.

Recently, very interesting results have been achieved with boron carbide NWs. The maximum Seebeck coefficient and the power were higher than in the bulk system [95]. Unfortunately, to date no experimental data of the thermal conductivity is available.

# 3    Measurement techniques

The measurement of all thermoelectric properties is a nontrivial task for which a wide range of measurement techniques has already been developed. This section will focus on different methods developed for measuring the thermal conductivity of nanostructures and how these can be combined with reliable electrical measurements. Thermal measurements can be performed either on single wires or on arrays, requiring very different measurement techniques. An overview of measurement techniques for single wires and for nanowire arrays will be given here and the advantages and disadvantages of each method will be briefly discussed. For more detailed reviews on measurement methods for thermoelectric properties of 1D nanostructures, the reader is encouraged to read more focused review [96–100].

## 3.1    Single nanowires

The measurement of single NWs is complicated by the low thermal conductance of these structures, requiring special care to be taken to remove any background thermal conduction. This has led to three distinct approaches for measuring the thermal conductance of NWs. Optical techniques employ the absorption of the optical excitation in the nanowire to ensure heating only the nanowire. Electronic techniques use a similar principle based on electronic self-heating, but this can only be used when the NW is electrically conductive. Finally a more direct method using heaters that are thermally isolated, except for a bridging through the nanowire has also been developed. All these methods will be described in more detail in the following sections.

### 3.1.1 Electronic techniques

#### 3.1.1.1 3-omega

If a wire is electrically conducting and reliable electrical contacts can be made to it, an electrical current can be sent through the wire in order to induce ohmic heating in the wire itself. Ohmic self-heating is used in the 3ω-method, which is well-established for bulk materials and thin films but has also been adapted for single NW measurements. A schematic of a typical setup used for these kind of measurements is shown in figure 8(a). In this method an AC heating current is applied with frequency ω and thus the power, and consequently the temperature and the resistance will vary with frequency 2ω. This modulation in the resistance, combined with the AC driving current leads to a voltage variation with frequency 3ω. By measuring voltage variations at a frequency 3ω the temperature variations caused by the heating power can be measured and can be used to model the thermal conductance. This method has been applied to measure the thermal conductivity of Bismuth[101], BiTe[102], $Bi_{0.8}Sb_{1.2}Te_{2.9}$[103], $Bi_{0.5}Se_{1.5}Te_3$[104], ZnO[105,106] and GaN[105] nanowires. The advantages of this system are the automatic inclusion of electronic contacts and a relatively easy processing. However the method is limited to samples with a relatively high electrical and thermal conductivity compared to the substrate and the evaluation of thermal contact resistances is a non-trivial problem.

### 3.1.2 Spectroscopy based method

Another method to locally measure the temperature of single NWs is based on spectroscopy. In this case, freely suspended NWs are heated by laser exposure and the determination of the local temperature by either photoluminescence [107] or micro-Raman spectroscopy [108]. In the case of photoluminescence measurements, the temperature is determined by the shift of the bandgap related emission, which is temperature dependent. In the case of Raman measurements, the lattice temperature is inferred from the shift of the phonon frequency, which is temperature dependent. In both cases, the laser beam is used simultaneously as a local heater and a local thermometer. The advantages of Raman over photoluminesce in this type of measurements is three-fold: i. Raman thermometry has a higher sensitivity; ii. can be applied to different type of materials, regardless of the character of the gap; iii Raman is a local technique, therefore offers a reliable spatial resolution. Therefore, in the following we will focus on Raman thermometry. By suspending a wire above a substrate and scanning with the laser over the length of the nanowire the thermal conductivity of the nanowire can be determined[109,110]. A schematic of the typical experimental setup used for this kind of experiment is shown in figure 8(b). If the measurement method is combined with EBL created electrical contacts all thermoelectric properties can be measured simultaneously[32]. Another possible configuration for Raman thermometry is provided by cantilevering the NWS[111]. The advantages of this method are the easy processing and the ability to create reliable electrical and thermal contacts. However, determining the thermal conductivity requires an accurate value for the absorbed laser power, which can be hard to obtain[110].

### 3.1.3 Thermal bridge method

A more direct way to measure the thermal conductivity of a single nanowire is using the microdevice proposed by Shi et. al.[112] consisting of two $SiN_x$ pads supported by long $SiN_x$ beams. On top of these pads a platinum meander is created as well as some electrodes for electrical contacts. A false colour SEM image of such a device is shown in figure 8(c). The platinum meander can be used both as a heater as well as a thermometer. When a wire is placed bridging the two pads the wire will transport the heat created in the heater to the thermometer on the other pad. Since the pads are thermally isolated, apart from mainly background radiation, thermal transport between the pads will be dominated by the nanowire. This allows for the measurement of thermal transport down to $1 nWm^{-1}K^{-1}$ with the sensitivity limited by electronic noise and thermal stability of the entire system. To further increase the sensitivity Wingert et. al. proposed an improvement of the electronics based on a Wheatstone bridge[113] improving the sensitivity by two orders of magnitude to $10 pWm^{-1}K^{-1}$. This makes this method the most sensitive method described here. However, difficult processing makes it difficult to reliably create electrical and thermal contacts. Since the influence of the thermal contact resistance increases with decreasing wire thermal resistance, this method works best for wires with a thermal conductance below a certain value[36]. This was also shown in a comparison between this method and the raman-based method for measuring the thermal conductivity of InSb nanowires with a diameter of 150-300 nm[110], where two different thermal conductivities were measured on similar wires, which was attributed to the thermal contact resistance.

### 3.1.4 T-type

A method similar to the microdevice method was recently proposed by Ma et. al.[114]. In this t-type method the wire is suspended on a thin Pt nanofilm that acts both as a heater and a thermometer. The other side of the wire is connected to the substrate which acts as a heat sink. A schematic of the device geometry is shown in figure 8(d). Since the Pt nanofilm has such a small thermal conductance, the wire acts as a substantial heatsink when placed on top of the nanofilm. Therefore the average temperature of the nanofilm as a function of power will decrease depending on the thermal conductivity of the wire. By comparing measurements with and without wire the thermal conductivity of the wire can be determined. This method sacrifices some sensitivity in favour of a considerably easier processing when compared to the microdevice method. However, the processing is still more challenging than the processing needed for the raman or 3ω method.

## 3.2 Arrays

### 3.2.1 Scanning Thermal Microscopy

In order to get spatial information of the thermal conductivity over the nanowire, besides optical scanning techniques, electrical scanning techniques can also be used. Scanning Thermal Microscopy (SThM) is a technique where a kelvin probe is combined with the 3ω method to get local heating and temperature measurements. A kelvin probe is an AFM probe with a conducting tip. This tip can be used simultaneously as heater and as thermometer through the 3ω method. In this way a NW can be heated up locally and the temperature can be measured as a function of heating power. This method is primarily being used to study NW arrays embedded in a polymer. By using the spatial resolution of the tip the contribution from the nanowire and the contribution from the embedding polymer can be separated resulting in the thermal conductivity of a single nanowire in the array. This technique has been used on a wide range of nanowires including $Bi_2Te_3$[115], Si[116], and SiGe[117]. Furthermore the technique has also been applied to $Sb_2Te_3$[118] and $In_3SbTe_2$[119] nanowires transferred on a substrate. The advantage of this measurement technique is the spatial resolution, which especially for embedded NWs helps to distinguish the different contributions to the thermal conductivity. The non-trivial problem of the contact resistance of the probe with the sample and the low sensitivity of the method are the main limitations of this method.

### *3.2.2 Thermoreflectance*

The thermal conductance of nanowire composites, *i.e.* arrays of NWs embedded in a hosting material can be measured using time-domain thermoreflectance (TDTR). These are pump-probe type of experiments with pulsed laser beams [120]. The pump beam induces a temperature rise at the sample surface, while the probe beam measures the evolution of the surface temperature through the change of the reflectivity with temperature, i.e. the thermoreflectance dR/dT, as a function of the relative delay time between pump and probe pulses. Persson *et al.* have performed TDTR measurements on composites of InAs NWs embedded in PMMA, obtaining a value of about $5.3 \pm 1.5$ W m$^{-1}$K$^{-1}$ for the thermal conductivity of the InAs NWs in agreement with theory and previous measurements of individual NWs [121]. Few years after, the thermal conductivity of composites of Si NWs with different surface roughness has been measured and a correlation between the thermal conductivity with rms roughness was found [122].

## 4 Overview of ZT values

Despite the significant effort devoted to developing measurement platforms that allow a complete determination of the thermoelectric properties of individual NWs, only relatively few experimental works have achieved this challenging goal. Table I lists the figure of merit and thermoelectric properties of the different types of semiconducting NWs, which have been experimentally determined. Beside the material system, also the diameter of the investigated NWs, whose thermoelectric properties are listed, is reported.

**Table I**: Thermoelectric properties of different types of NWs.

| NW material | Diameter (nm) | Thermal conductivity (Wm$^{-1}$K$^{-1}$) | Electrical conductivity ($\Omega^{-1}$cm$^{-1}$) | Power factor (10$^{-4}$ Wm$^{-1}$K$^{-2}$) | ZT | REF |
|---|---|---|---|---|---|---|
| Si | 50 | 1.6 | 588 | 33 | 0.6 | [5] |
| Si | 20 | 2 | 571 | 100 | 1.0 | [6] |
| InSb | 177 | 6 | 100 | 1.0 | 0.01 | [31] |
| InSb | 150 | 17 | 140 | 15 | 0.025 | [32] |
| InAs | 125 | 2.6 | 20 | 46 | 5.4·10$^{-4}$ | [35] |
| Bi$_2$Te$_3$ |  | 1.2 | 500 | 0.072 | 0.2 | [37] |
| Bi$_{0.39}$Te$_{0.61}$ | 270 | 0.9 | 1300 | 2.6 | 0.08 | [102] |
| Bi$_{1-x}$Te$_x$ | 52 | 3 | 1900 | 7.5 | 0.1 | [38] |
| Bi$_{1-x}$Te$_x$ | 81 | 1.0 | 700 | 0.66 | 0.02 | [123] |
| Bi$_2$S$_3$ | 153.4 | 2.6 | 0.64 | 0.018 | 1.76·10$^{-4}$ | [114] |
| Bi$_{0.8}$Sb$_{1.2}$Te$_{2.9}$ | 750 | 0.72 | 780 | 24 | 1.0 | [103] |
| PbSe | 50-100 | 0.8 | 0.25 | 3.2 | 0.12 | [44] |
| SnTe | 508 | 4.2 | 588 | 5.3 | 0.0379 | [94] |
| SiGe | 26 | 1.2 | 600 | 12 | 0.46 | [53] |
| SiGe | 260 | 2 | 400 | 12 | 0.18 | [52] |

| Bi/Te C/S | 456 | 2.6 | 3910 | 44 | 0.5 | [10] |

Si and SiGe nanowires, are still among the most promising candidates for thermoelectric applications, while III-V semiconductor NWs (*i.e.* InAs and InSb) exhibit a figure of merit below the calculated one. However, to date the ZT of InAs and InSb nanowires was measured only on relatively large diameters due to experimental challenges. This suggests that a crucial aspect in the field is still the improvement of experimental methods for the accurate determination of both thermal and electrical properties at the nanoscale. The heavy elements NWs provided reasonable but not astonishing figure of merit. However, two ZT values stand out in Table I: those related to $Bi_{0.8}Sb_{1.2}Te_{2.9}$ and Bi/Te core/shell NWs. These results indicate in our opinion an interesting pathway in terms of material design: the combination of more ingredients like, *e.g.*, heterostructuring and alloying.

## 5 Conclusions

NWs represent the ideal material system whose thermoelectric properties can be tuned by design. On top of the "conventional" (bulk-like) approaches for the enhancement of the thermoelectric properties, *e.g.* alloying, other characteristics of the NWs can be exploited to engineer their properties, like the geometry, growth direction, heterostructuring, and crystal phase. The size and surface-to-volume ratio will modulate the boundary scattering. Considering the NW-based heterostructures, besides offering two different types of heterostructure geometry, *i.e.* axial and radial, NWs provide a larger portfolio of materials which can be combined together thanks to the radial relaxation of the strain. Furthermore, a new type of heterostructures, namely crystal phase heterostructures, can be realized with effect on the thermoelectric properties. Still, the experimental assessment of all the thermoelectric properties of individual or arrays of NWs represent a challenge for the field.

## Acknowledgments

We acknowledge financial support from the NWO through the VENI grant (680-47-443) and the ERC Starting Grant PHONUIT (Grant 756365).

# Figures

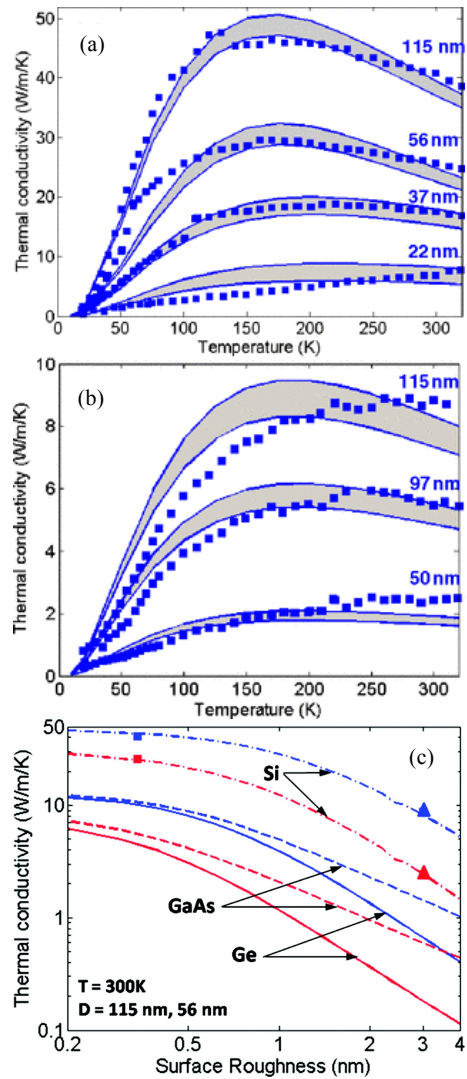

**Figure 1**: (a) Thermal conductivity of smooth VLS Si NW. Shaded areas are theoretical predictions with roughness rms $\Delta = 1-3$ Å, blue squares are taken from [124]. (b) Thermal conductivity of rough EE Si NW ($\Delta = 3-3.25$ nm). Squares are taken from [5]. Simulation and experimental data are compared at similar cross sections. L = 6nm. Reprinted (adapted) with permission from [19], copyright 2009 American Physical Society. (c) Effect of roughness rms on thermal conductivity of Si, Ge, and GaAs NW with diameters (D) of 115 nm (upper curve) and 56 nm (lower curve). L = 22 Å throughout. Reprinted (adapted) with permission from [20], copyright 2010 American Chemical Society.

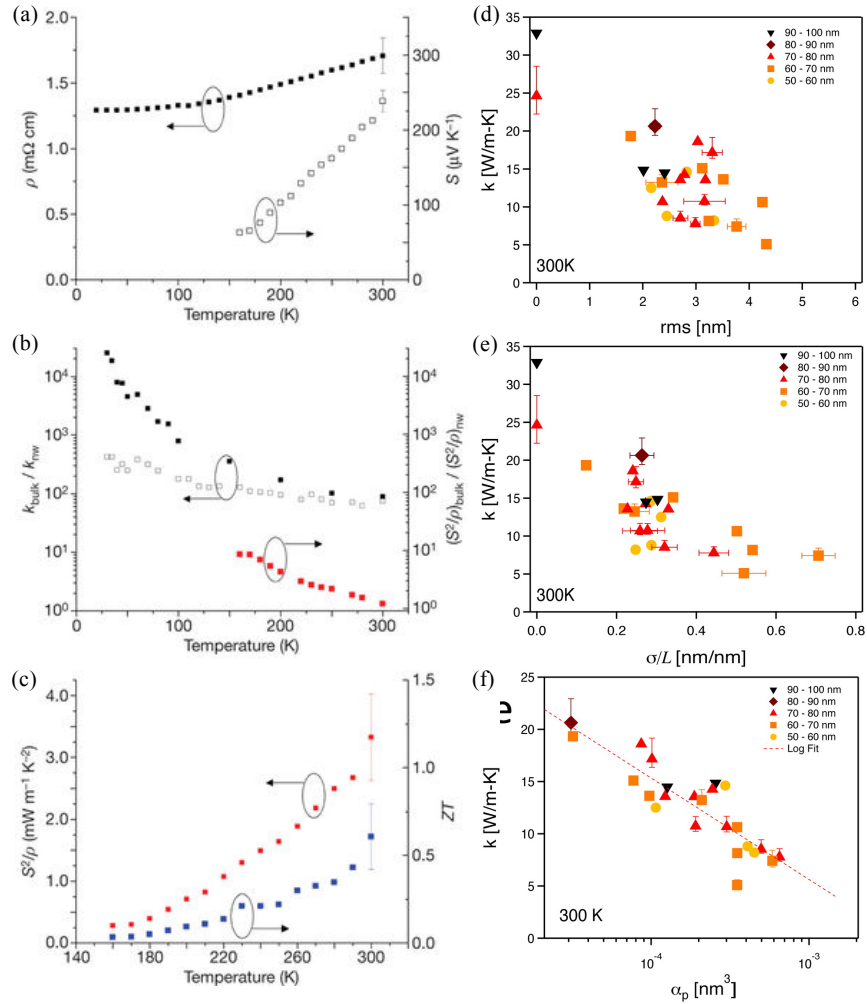

**Figure 2**: (a) S (open squares) and ρ (solid squares) of the highly doped electroless etched 48nm nanowire. (b) Ratio of bulk Si k to that of a highly doped electroless etched Si nanowire 50nm in diameter $k_{bulk}/k_{nw}$ for intrinsic bulk Si (solid squares, [125]) and highly doped bulk Si (open squares, data adapted from [126]). Red squares show the ratio of the power factor of optimally doped bulk Si relative to the nanowire power factor as a function of temperature. (c) Single nanowire power factor (red squares) and calculated ZT (blue squares) using the measured k of the 52 nm nanowire. By propagation of uncertainty from the r and S measurements, the error bars are 21% for the power factor and 31% for ZT (assuming negligible temperature uncertainty). Reprinted (adapted) with permission from [5], copyright 2008 Springer Nature. (d) Thermal conductivity as a function of rms with different range of diameter. (e) Thermal conductivity as a function of ratio between rms and correlation length σ/L. (f) Thermal conductivity as a function of $α_p$ (plotted on a log scale), which is a parameter related to the broadband roughness amplitude parameter of the nanowire surface and increases with the roughness. Reprinted (adapted) with permission from [21], copyright 2012 American Chemical Society.

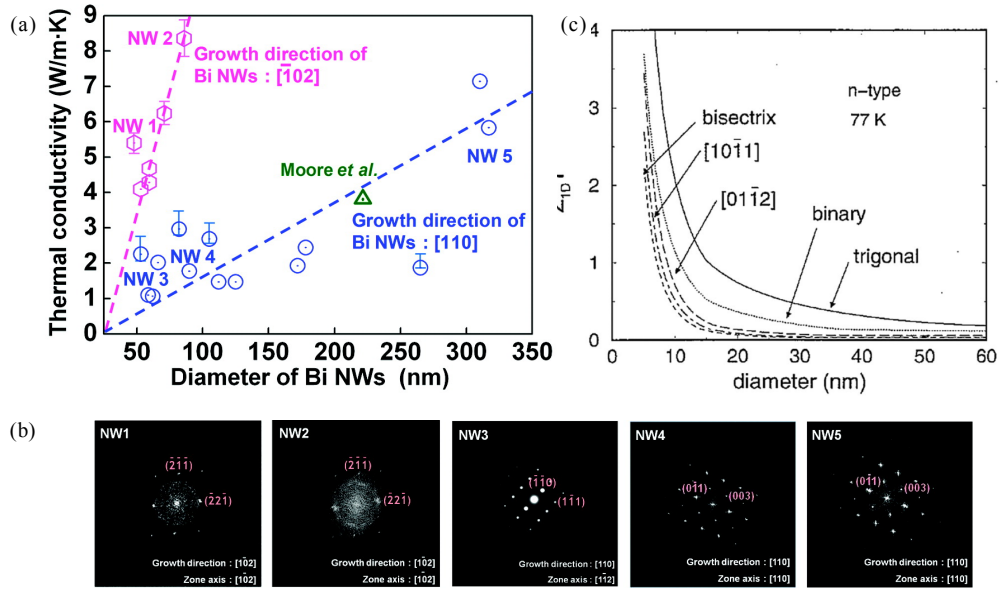

**Figure 3**: (a) Diameter-dependent thermal conductivity of Bi nanowires with growth directions of $[\bar{1}02]$ (pink hexagon) and [110] (blue circle), respectively, at 300 K. The dashed lines represent the linear fit of the measured thermal conductivity for each growth direction of Bi nanowires. The thermal conductivity (green triangle) of Bi nanowire perpendicular to the trigonal axis measured by Moore et. al. is taken from [28]. (b) SAED patterns of the thermal conductivity-measured Bi nanowires. While the ED pattern shows NW 1 and NW 2 were grown along the direction of $[\bar{1}02]$, NW 3, NW 4, and NW 5 were grown along the direction of [110]. Reprinted (adapted) with permission from [29], copyright 2011 American Chemical Society. (c) Calculated $Z_{1D}T$ at 77 K as a function of diameter for n-type Bi nanowires oriented along different directions. Reprinted (adapted) with permission from [27], copyright 2000 American Physical Society.

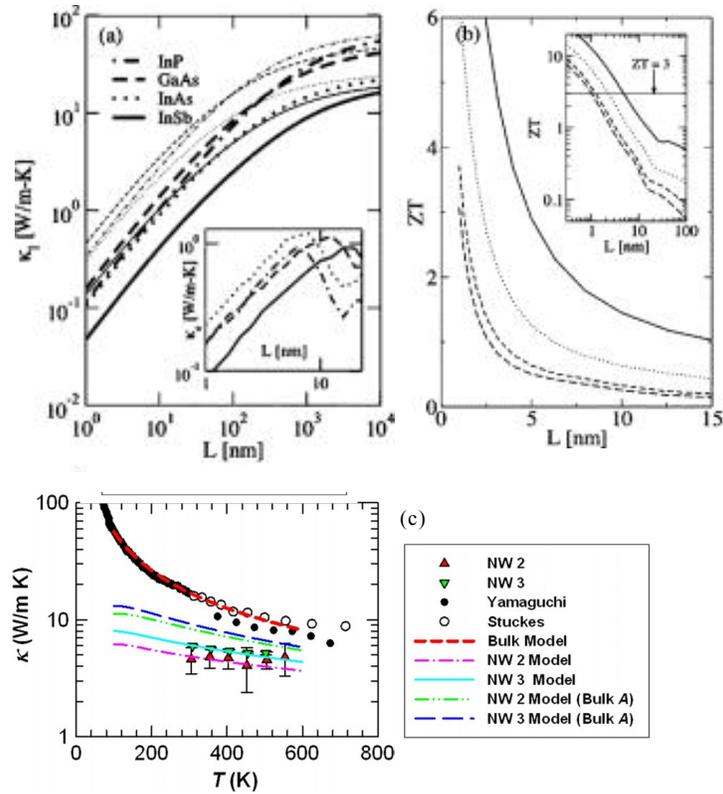

**Figure 4**: a) Theoretically calculated lattice thermal conductivity versus thickness, for InSb, InAs, GaAs, and InP nanowires, for diffusive (thick lines) and partially specular (thin lines) phonon boundary scattering. Inset: electronic contribution to the thermal conductivity. b) Calculated ZT versus thickness (thick: diffusive boundary scattering; thin: partially specular). Inset: ZT with diffusive boundary scattering, in logarithmic scale. Reprinted (adapted) with permission from [30], copyright 2004 AIP Publishing LLC. (c) Thermal conductivity as a function of temperature for two different InSb NWs. Bulk InSb data found in the literature were plotted together for comparison [127,128]. Dashed and solid lines are calculation results using a modified Callaway model for bulk InSb and NWs with a diffuse surface and the same diameter as the two measured NWs and different impurity scattering parameters. Reprinted (adapted) with permission from [31], copyright 2010 Macmillan Publishers Limited.

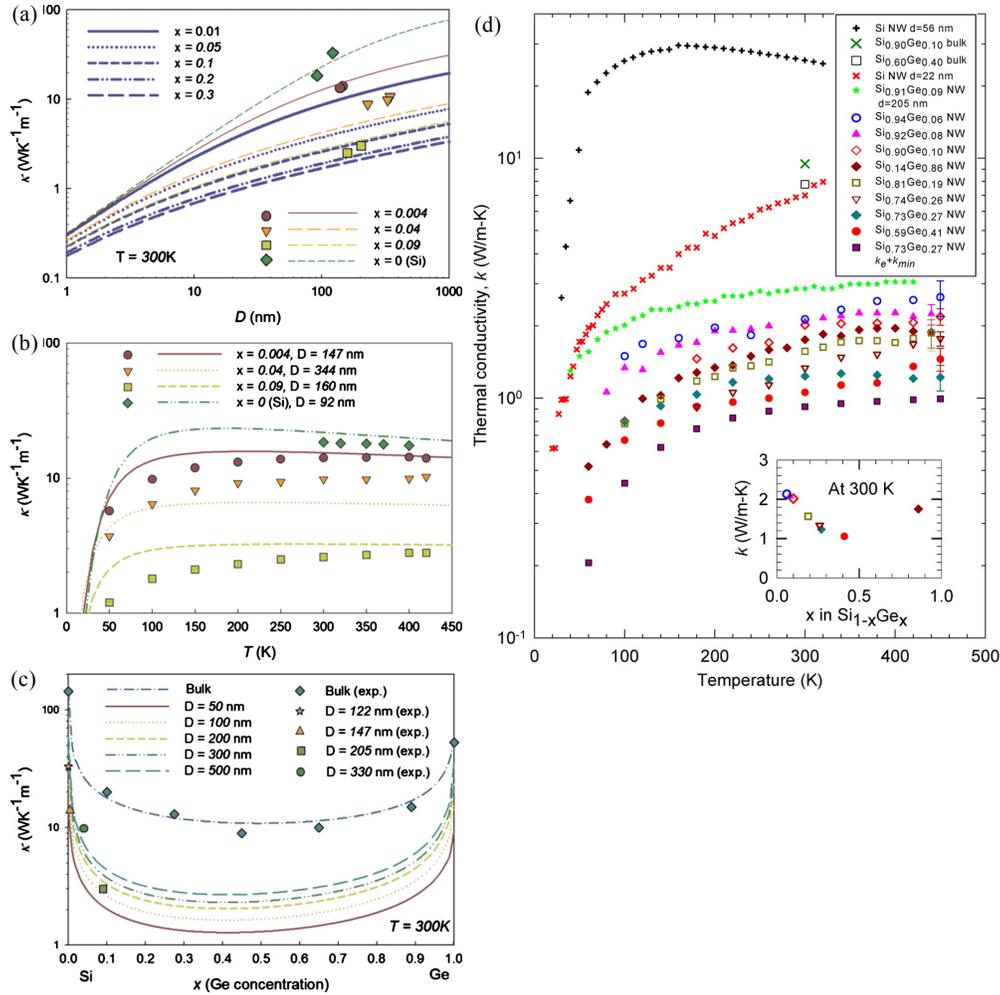

**Figure 5**: Thermal conductivity as a function of NW diameter a) and temperature b) for $Si_{1-x}Ge_x$ NWs with different germanium concentrations. The lines show calculation, while the symbols represent experimental results from [129]. c) Thermal conductivity versus germanium concentration for $Si_{1-x}Ge_x$ bulk alloy and NWs with different diameters. The lines show calculation results, and the symbols represent experimental data from [129,130]Reprinted (adapted) with permission from [51], copyright 2010 AIP Publishing LLC. d) The thermal conductivities of the SiGe NWs, plotted with those of the previously reported samples including a 56 nm diameter Si nanowire[124], a $Si_{0.9}Ge_{0.1}$ bulk[131], a $Si_{0.6}Ge_{0.4}$ bulk[131], a $Si_{0.91}Ge_{0.09}$ nanowire[129], and the calculated summation of electronic ($k_e$) and minimum ($k_{min}$) thermal conductivities for the $Si_{0.74}Ge_{0.26}$ NW. The inset shows the thermal conductivities at 300 K of the SiGe nanowires that were measured in this study. Reprinted (adapted) with permission from [53], copyright 2012 American Chemical Society.

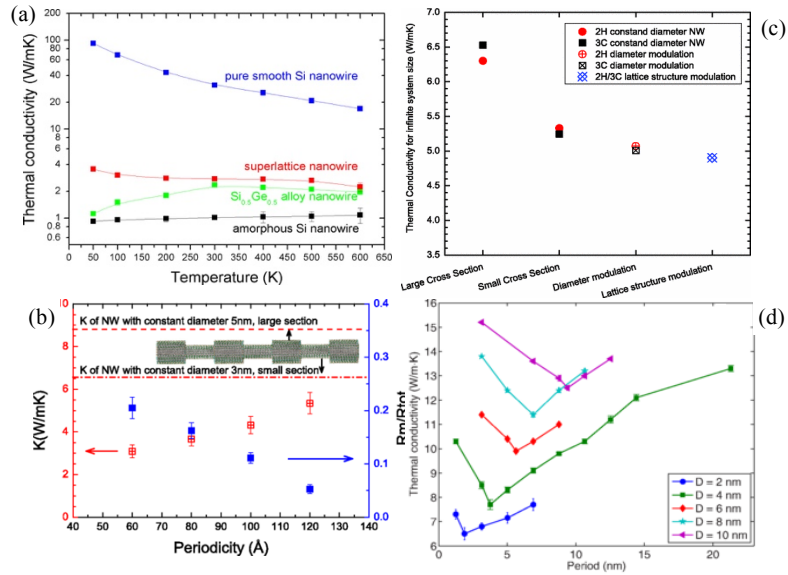

**Figure 6**: a) Temperature dependence of the thermal conductivity of Si/ Ge superlattice nanowires in comparison with pure smooth Si nanowires, $Si_{0.5}Ge_{0.5}$ alloy nanowires, and fully amorphous Si nanowires. All wires have the same cross section of $3.07 \times 3.07$ nm$^2$ and the same length of 278 nm. Reprinted (adapted) with permission from [65], copyright 2012 American Chemical Society. b) Evolution of thermal conductivity and of the ratio Rm/Rtot as a function of the periodicity P of the diameter modulations. Reprinted (adapted) with permission from [68], copyright 2015 American Physical Society. c) Thermal conductivity for infinite system size for the constant cross section 2H and 3C monotype NWs, the diameter modulated 2H and 3C nanowires, and the lattice modulated 2H/3C nanowires. Reprinted (adapted) with permission from [69], copyright 2013 American Physical Society d) Thermal conductivities of the twinning SL NWs as a function of the period for different diameters at 300 K. A minimum thermal conductivity appears at different period lengths for different diameters. Reprinted (adapted) with permission from [70], copyright 2014 American Physical Society.

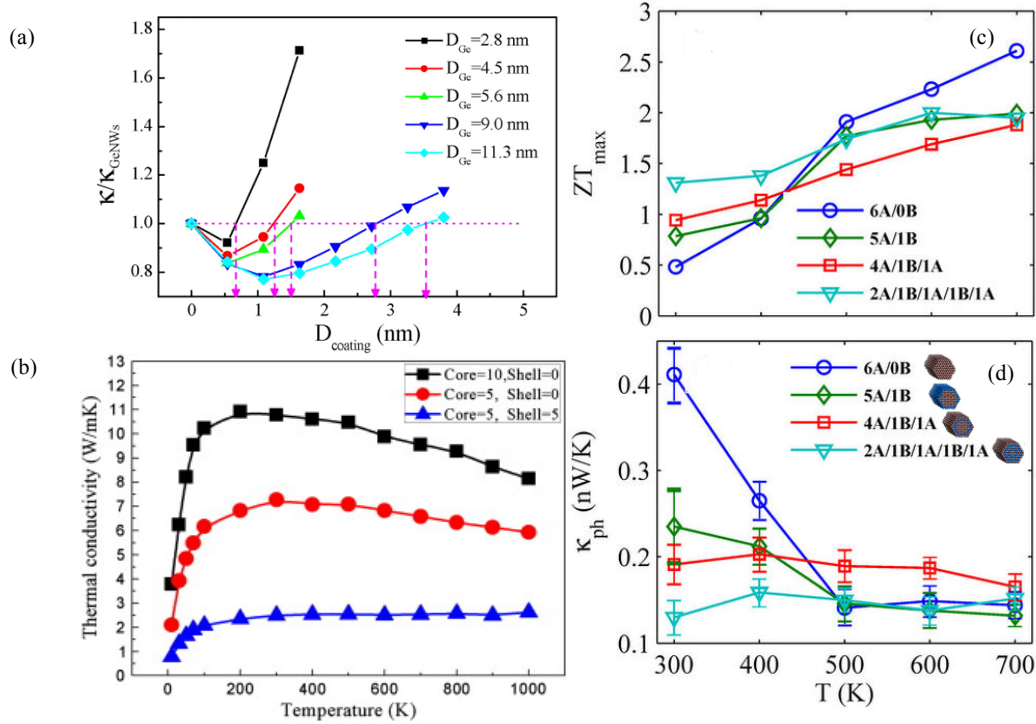

**Figure 7**: a) Normalized thermal conductivity versus coating thickness for different $D_{Ge}$. Thermal conductivity of GeNWs at each DGe is used as reference. The dashed arrows point the critical coating thickness when thermal conductivity of Ge/Si core−shell NWs ($\kappa_{core-shell}$) is equal to that of GeNWs ($\kappa_{GeNWs}$). The dashed line is drawn to guide the eye. Reprinted (adapted) with permission from [85], copyright 2012 American Chemical Society b) The thermal conductivity curves of crystalline SiNWs and Si/a-Si core–shell NWs as a function of temperature. Reprinted (adapted) with permission from [89], copyright 2014 Elsevier. c)Temperature dependence of ZT for a uniform GaSb NW (blue cirkels), a GaSb/InAs core-shell NW (green diamonds), and a 2 (red squares) and 4 shell (light blue triangles) MCW Reprinted (adapted) with permission from [84]copyright 2014 Springer Nature. d) Temperature dependence of the thermal conductivity ($\kappa_{ph}$) for a uniform GaSb NW (blue cirkels), a GaSb/InAs core-shell NW (green diamonds), and a 2 (red squares) and 4 shell (light blue triangles) MCW Reprinted (adapted) with permission from [84], copyright 2014 Springer Nature.

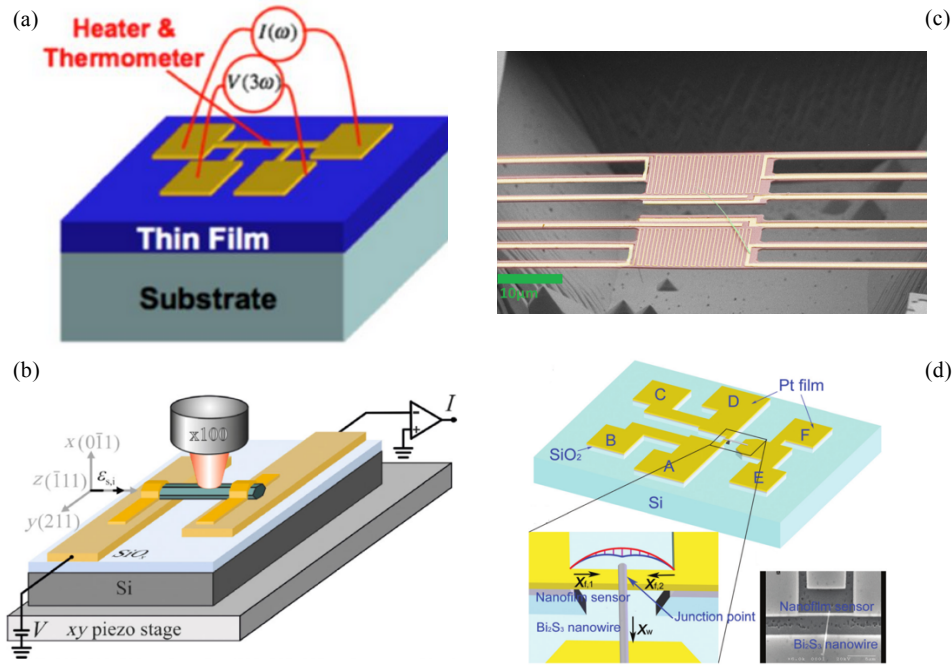

**Figure 8**: a) Schematic diagram of a typical 3 ω measurement set-up b) Schematic of the device and scattering geometry used for Raman spectroscopy. The InSb NW is suspended between two gold stripes evaporated on a SiO2/Si substrate and clamped with two top contacts. A voltage is applied at one end of the NW, and the current is measured through a low-noise current amplifier. The Raman scattering configuration is also shown together with the NW geometry. The entire device is located on a xy piezo stage. Reprinted (adapted) with permission from [32], copyright 2015 Springer Nature 1 c) False color SEM picture of the device used for measuring the thermal conductivity of nanostructures consisting of 1mm long SiNx bridges supporting two 4 micron separated membranes. On both membranes a platinum meander functioning as a heater or resistance thermometer is fabricated. A nanowire is placed between the membranes, which results in a heat conductance Gx between the two membranes. d) Schematic of the T-type method Reprinted (adapted) with permission from [114], copyright 2016 Royal Society of Chemistry